\documentclass[11pt]{article}
\usepackage{amsmath}
\usepackage{amsfonts}
\usepackage{amssymb}
\usepackage{graphicx}
\usepackage{subfigure}
cm \headheight=0. mm
\begin{document}
\begin{center}
\LARGE {Stability of Ghost Dark Energy in  CBD  Model of Gravity }
\end{center}
\begin{center}
{\bf Kh. Saaidi\footnote{ksaaidi@uok.ac.ir,  {\rm or},  ksaaidi@phys.ksu.edu}}\\

{\it Department of Physics, Faculty of Science, University of
Kurdistan,  Sanandaj, Iran}\\
{ \it Department of Physics, Kansas State University,116 Cardwell Hall, Manhattan, KS 66506, USA.}\\
%{\it $^c$Faculty of Science,  Islamic Azad University Sanandaj Branch, Sanandaj, Iran }
\end{center}
 \vskip 1cm
\begin{center}
{\bf{Abstract}}
\end{center}
We study the stability of the ghost dark energy model versus perturbation. Since this kind of dark energy is instable in Einsteinian general relativity theory, then we study a new type of Brans-Dicke theory which has non-minimal coupling with matter which is called chameleon Brans-Dicke (CBD) model of gravity. Due to this coupling the equation of conservation energy is modified. For considering the stability of the model we use the adiabatic squared sound speed, $c_s^2$, whose sign of it determines the stability of the model in which   for $c_s^2 >0 $ the model is stable and for $c_s^2 <0 $  the model is instable. However,  we study the interacting and non-interacting version of chameleon  Brans-Dicke ghost dark energy (CBDGDE) with cold dark matter in non flat FLRW metric. We show that in all cases of investigation the model is stable with a suitable choice of  parameters.\\

\newpage
\def\br{\biggr}
\def\bl{\biggl}
\def\Br{\Biggr}
\def\Bl{\Biggl}
\def\be{\begin{equation}}
 \def\ee{\end{equation}}
\def\bea{\begin{eqnarray}}
\def\eea{\end{eqnarray}}
\def\f{\frac}
\def\n{\nonumber}
\def\l{\label}
\def\ov{\over}
\def\om{\omega}
\def\Om{\Omega}
\def\p{\phi}
\def\R{\rho}
\def\va{\beta}

\newpage

%=================================================================
%============== Introduction =====================================
%=================================================================

\newpage

\section{Introductions}
The cosmological and astrophysical observations such as type Ia supernovae data{\cite{1}}, Wilkinson Microwave
Anisotropic Probe (WMAP) {\cite{2}}, X-ray {\cite{3}}, large scale
structure {\cite{4}} and ete,  indicate that
our universe is in accelerating  expansion phase. People
have introduced  an energy component of the universe, called dark energy to describe this acceleration. The simplest model of DE  is a tiny positive time-independent cosmological
constant, $\Lambda$, for which has the equation of state
$\omega=-1$ {\cite{7, 8}}. Plenty of other DE models
have also been proposed for  explain the acceleration
expansion either by introducing new degree(s) of freedom or by modifying gravity \cite{9, 10, 11, 12, 13, 14, 15, 16, 18}.\\

In recent years there has been a new attention to the so called "scalar-tensor gravity".  The scalar-tensor models
include  a  scalar field, $\phi$, whit non-minimal coupling
to  the geometry in the gravitational action, which has been introduced by
Brans and Dicke (BD) {\cite{20}}. They proposed a scalar degree of freedom to incorporate the Mach's principle into general relativity.
The  mechanism that creates a  non-minimal scalar field
coupling to the geometry,  can also lead to a
coupling between the scalar and matter field. Therefore authors {\cite{24}}  introduced a scalar
field which it has a coupling to matter with  order unity strength, named
chameleon mechanism.  Indeed, the chameleon
proposal provides a way to generating  an effective mass for  a light scalar
field via the  field self interaction, and the  interaction between  matter
 and scalar field.\\

%================================================================
Recently, a new kind of  dark energy model, so called "ghost dark energy" has been investigated {\cite{31, 32}}.  Originally, the Veneziano ghost was introduced as a solution to $U(1)$ problem in low energy  effective theory of QDF \cite{33}. The ghosts  make a small
energy density contribution to the vacuum energy due to the off-set of the cancelation of
their contribution in curved space or time-dependent background. The authors of \cite{32} have clarified the decoupling of the QCD vector ghost to the vacuum energy density in the Rindler space-time.  They  have fund that it gives the vacuum energy density proportional to Hubble parameter, $H\Lambda^3_{QCD}$  of the right magnitude $\sim (10^{-3} ev)^4$, where $H$ is the Hubble parameter and $\Lambda_{QCD}$is QCD mass scale. The authors of  \cite{32} have climbed  that   in the ghost model of dark energy, one needs not to introduce any new degree of freedom or modify gravity and it is totally embedded in standard model of gravity. But according to results of \cite{35} the GDE model in the Einsteinian theory of gravity   has a behavior like a cosmological constant at the late time ($\omega =-1$) and the equation of state can never  cross $-1$, and this is similar to the behavior of quintessence. Also, they have shown that the adiabatic squared sound speed of the GDE model in context of standard model of cosmology  is negative and then this model can not be stable. Therefore, in this work  we want to consider the stability of GDE model in context of BD model which has a non-minimal coupling with matter. The fundamental  key quantity for studying  the stability of a model is the squared adiabatic sound speed which is obtained a small perturbation  in the back ground energy density \cite{20s}.
The sign of $c^2_s$ plays a crucial role in determining the stability of the
background evolution. If $c^2_s <0 $, it means that the model is  classically instable against  perturbation.
This issue has already been investigated for some DE models  such as chaplygin
gas and tachyon DE \cite{21s}, holographic DE \cite{22s}, agegraphic model of DE \cite{23s}.

In this work, we study  the  ghost model of dark energy in the context of chameleon  Brans-Dicke  model of gravity.   We investigate  the cosmological evolution of  our
model with/without interaction between  DE  and  cold dark matter. We analytically and
numerically compute some quantities such as scale factor $a$, EoS parameter of dark energy $\om_d$, deceleration parameter $q$,  fraction of dark energy $\Om_d$,    squared adiabatic speed of sound $c^2_s$  and so on.\\

This work is organized in four sections, of which this introduction is the first. In section two, the action is introduced and the field equation, the scalar field equation of motion, the modified conservation of density energy  are obtained. In section 3 the interacting and non interacting GDE model of  CBD theory is investigated in non flat FRW space-time, and section 4 is the summarize of our results.

%==========================================================================
%======================== Section 2 () =============================
%=========================================================================

\section{General Framework}
For our investigation, we consider the chameleon-Brans-Dicke action
\begin{equation}\label{1a}
S=\int d^4x\sqrt{-g} \left( \phi R -
\frac{\omega}{\phi}\partial_{\mu}\phi \partial^{\mu} \phi
 +2f(\phi)\mathcal{L}_m \right),
\end{equation}
where $g$ is the metric determinant, $R$ is the Ricci scalar
constructed from the metric $g_{\mu \nu}$, and $\phi R$ has been
replaced with the Einstein-Hilbert term is such a way that
$G_{eff}^{-1}=16\pi\phi$, $\phi$ is the
chameleon-Brans-Dicke scalar field, $\omega$ is the dimenssionless
Brans-Dicke constant.  The last term
on the right hand side of (\ref{1a}),
$f(\phi)\mathcal{L}_m$, indicates non-minimal coupling between the
scalar field and matter. One can obtain the gravitational field equation by
taking variation of the action (\ref{1a}) with respect to the
metric $g_{\mu \nu}$
\begin{equation}\label {2a}
R_{\mu \nu}-\frac{1}{2}g_{\mu \nu}R={f(\phi)\ov \phi}  T_{\mu\nu}+T^{\phi}_{\mu\nu},
\end{equation}
where $T^{\phi}_{\mu \nu}$ indicates the scalar field  energy-momentum tensor  which is
\begin{equation}\l{3a}
T^{\phi}_{\mu\nu} =\frac{\omega}{\phi^2}\bigg[\partial_{\mu} \phi \partial_{\nu} \phi
- \frac{1}{2}g_{\mu \nu}(\partial_{\alpha}\phi)^2\bigg]
 +{1\ov \p} \bigg[\partial_{\mu} \partial_{\nu} -g_{\mu \nu}\Box\bigg]\phi ,
\end{equation}
here  $\Box$ is the four dimensional d'Alambert operator, and
$T_{\mu \nu}$ indicates the matter\footnote{ In fact matter stress-energy tensor consists of all perfect fluids stress-energy tensor, namely $T_{\mu\nu} = T^{(b)}_{\mu\nu}+T^{(cdm)}_{\mu\nu}+T^{(r)}_{\mu\nu}+T^{(d)}_{\mu\nu}$. Here the subscript $b$, $cdm$, $r$ and $d$ indicate baryonic matter, cold dark matter, radiation and dark energy respectively. Indeed in this work, the same as others, we assume that dark energy has an averaged bahavior like perfect fluid.}   energy-momentum tensor  which is defined
\be\l{3a'}
T_{\mu\nu}=  -{2\ov \sqrt{-g}}{\delta (\sqrt{-g}{\cal L}_m)\ov\delta g^{\mu\nu}},
\ee
  and represented
by
\begin{equation}\l{4a}
T_{\mu\nu} = (\rho_t+p_t)u_{\mu}u_{\nu} + p_tg_{\mu\nu}.
\end{equation}
Where $u_{\mu}$ is the four-vector velocity of the fluid satisfying $u_{\mu}u^{\mu}=-1$,   $\rho_t$ and $p_t$
are respectively the total energy density and total isotropic pressure of the barotropic perfect fluids which have filled the universe.
 Taking
variation of action with respect to  scalar field $\phi$ gives
us the Klein- Gordon equation for the   scalar field  as
\begin{equation}\label {5a}
(2\omega+3)\Box \phi=\big[f(\p) T - 2\p f'(\p) {\cal L}_m\big],
\end{equation}
where, ${\cal L}_m$ is the Lagrangian of the matter and  $T$ is the trace of matter stress-energy  tensor.  The Bianchi identities, together with the identity
$(\square\nabla_a - \nabla_a\square  ) V_c = R_{ab}\nabla^aV_c$, imply the non-(covariant)
conservation law
\be\label{7'}
\nabla_aT^{ab} = \big[g^{ab}{\cal L}_m + T^{ab}\big]{\nabla_af\ov f},
\ee
and, as expected, in the limit $f(\p)$ = constant, one recovers the
conservation law $\nabla_aT^{ab} = 0$.\\

Our aim in this work is to consider the ghost model of dark energy in context of   CBD model in the Friedmann-Robertson-Walker (FRW) Universe which is described by the following line element
\begin{equation}\l{6a}
ds^2 = dt^2 - a(t)^2 {\biggr(}\frac{dr^2}{1-kr^2}+ r^2d\Omega^2 { \biggl )},
\end{equation}
where $a(t)$ is the scale factor, and $k$ is the curvature parameter with $k = -1, 0, 1 $ corresponding to open, flat, and
closed Universes, respectively. By making use ({\ref{2a}}), (\ref{3a}), (\ref{4a}), (\ref{5a}) and (\ref{6a}) one can arrive at
\begin{eqnarray}\label {7a}
3H^2 + {k \ov a^2}&=& {f(\p) \ov \p}\rho_t+\frac{\omega}{2}(\frac{\dot{\phi}}{\phi})^2-3H(\frac{\dot{\phi}}{\phi}) \\
2\dot{H}+3H^2+\frac{k}{a^{2}}&=&-{f(\p) \ov \p}{p_t}
-\frac{\omega}{2}\frac{\dot{\phi}^{2}}{\phi^{2}}-2H(\frac{\dot{\phi}}{\phi})
-\frac{\ddot{\phi}}{\phi},\l{8}\\
 \ddot{\phi}+3H\dot{\phi}&=&\frac{1}{(2\omega+3)}\big[f(\p) T - 2\p f'(\p) {\cal L}_m\big],\l{9}
\end{eqnarray}
where $H={\dot{a} / a}$ is the Hubble parameter. In the above equations,  the
EoS  parameter of the baryonic and dark matter is $p_t = \om_t \rho_t$.

In \cite{Hok}, it is shown that   a "natural choice" for the matter
Lagrangian density for perfect fluids which based on (\ref{3a'})  can  give us the stress-energy tensor, (\ref{4a}),    is $ {\cal L}_m = p_t$,  where $p_t$ is the pressure. However, although $ {\cal L}_m = p_t$ does indeed reproduce the perfect fluid
equation of state, it is not unique. For example,  other choices are  $ {\cal L}_m = -\rho_t$ or $-n_t a_t$, where $\rho_t$ is the energy
density,  $n_t$ is the total particle number
density, and $a_t$ is the total  physical free energy defined as
$a_t = \rho_t/n_t - {\cal T} s$, with ${\cal T }$ being the fluid temperature and $s$
the entropy per particle \cite{Hok}. Therefore one may  introduce the matter Lagrangian as\footnote{As mentioned before, the choice of ${\cal L}_m$ is not unique but only for  simplicity we have chosen it as (\ref{10'}).   }
\be\l{10'}
{\cal L}_m = -{1\ov 4}\rho_t +{3\ov 4} p_t,
 \ee
 which can give us the stress-energy tensor, (\ref{4a}), based on (\ref{3a'}). So using (\ref{6a})  we can rewrite  the $tt$ component of (\ref{7'}) as follows
\be\l{9'}
\rho_t +3H(1+\om_t)\rho_t = -{3\ov 4}{\dot{f} \ov f}(1 + \omega_t)\rho_t.
\ee

 It is seen that there is an addition parameter in evolution equation of model.
 So according \cite{36} we shall assume that the  scalar field can be introduced as a power law of the scale factor  as
\begin{equation}\l{11}
\phi =N({a(t) \ov a_0})^{\xi} = N a^{\xi},
\end{equation}
here $a_0$ is the scale factor at the present time, $a_0 =1$, and $N= $ constant.
 In fact there is no compelling reason for
this choice. However, it has been shown that for small $\xi$ it leads to consistent results and the product $|\xi|\omega$ results of
order unity \cite{36}. According to the solar system experiments the magnitude of   $\omega$ is more than  $ 40000$ ($\omega > 40000$) \cite{37, 38} and this means  that  the value of $\xi$ has to very small $\sim 10^{-4}$.
Using the latest WMAP and SDSS data, the observational constraints on BD model in a flat Universe with cosmological constant and cold DM  is obtained \cite{39}. They found that within
$2\sigma$ range, the value of $\omega$ satisfies $\omega < -120.0$ or $\omega > 97.8$. They also obtained the constraint on the rate of change
of $G$ at present as $-5.53 \times 10^{-20} s^{-1}<{\dot{G}/ G} <3.32  \times 10^{-20} s^{-1}$,
at $2\sigma$ confidence level. So in our case with assumption (\ref{11}) we get $-5.53 \times 10^{-20} s^{-1}<{\dot{G} / G}= {\dot{\phi} / \phi}=\xi H <3.32  \times 10^{-20} s^{-1}.$
This relation can be used to put an upper and lower  bound on $\xi$.  Assuming the present value of the Hubble parameter to be
$H_0\simeq 2.11\times 10^{-18} s^{-1}$, we obtain
\begin{equation}\l{12}
-0.026< \xi <0.015.
\end{equation}

For getting a better insight, we continue our work based on  a power law form for $f(\p)$ as $f(\p)= f_0\p^s$,
 where $f_0$ and here $s=1$\footnote{The results of this study based on $s\neq 1 $ is not very different to the our investigation. } are constant. Also
 we assume there are only two components GDE and CDM in the  Universe\footnote{ Since we are interesting to study the our model at the late time and in this stage  dark energy is dominant, then for simplicity we absorb other components of perfect fluid(baryonic matter and radiation) in cold dark matter part.}.  Therefore $p_t = p_m + p_d = p_d$ and then $p_t = \om_t\rho_t = \om_d \rho_d$, where $\rho_t = \rho_d+\rho_m$. Hence from (10) and using (11), we have
\begin{eqnarray}\l{13}
\dot{\rho}_m + 3H(1 + {\xi \ov 4}) \rho_m &=& 0, \\
\dot{\rho}_d + 3H(1 +   \om_d)(1 + {\xi \ov 4}) \rho_d &=& 0.\l{14}
\end{eqnarray}
 One can interpret the interaction term in (\ref{9'}) and also in (\ref{13})  as
$p_m = \xi \rho_m/4$. This means that for $\xi<0$, the interaction between chameleon scalar filed and matter fluid  can create a negative pressure.
Therefore we expect this process help to describe the positive accelerating expansion of Universe.
\section{GDE in the context of CBD model of gravity}
In this section
we  consider the GDE in a non flat space-time of FRW Universe in chameleon BD model of gravity. The ghost energy density is proportional to the Hubble parameter \cite{32}.
\begin{equation}\label{15}
\rho_d = \alpha H,
\end{equation}
where $\alpha$ is a constant. According to the results of \cite{32}, $\alpha\sim \Lambda^3_{QCD}$, where $\Lambda \sim 10^2$MeV is QCD mass scale. Therefore the value of dark energy at the present time, with    $H \sim 10^{-39}$MeV is about $(3\times 10^{-3 }eV)^4$. This value is in an excellent agreement with observed DE density \cite{32}. As mentioned in the Introduction this model of dark energy  in the context of standard model of cosmology is not stable and this means that GDE model in standard model of gravity has shortcoming and need to study in another model of gravity.

\subsection{Non interacting GDE }

In this subsection we investigate the GDE model in the CBD framework in a non flat FRW space time and we will obtain the EoS parameter,  deceleration parameter, the evaluation of fractional energy density and the adiabatic squared sound speed  of  GDE.
Taking the time derivative of equation (\ref{7a}),  using relation  (\ref{11}) and  the  continuity equations (\ref{14}), we find
\begin{eqnarray}\l{16}
{\dot{H} \over H^2}&=&-3\beta(1+\om_d),\n\\
 &=&   -{1 \over 2\theta}{\biggr \{}3\beta\theta-(2-3\beta)\Omega_k{\biggl  \}}-{3 \over 2\theta}\omega_d \Omega_{d},
\end{eqnarray}
  where $\theta = 1- \xi({\omega\xi / 6}-1)$,  $\beta = 1 + s\xi/4$. One can obtain the EoS parameter of  GDE model of  chameleon BD  as
 \begin{equation}\l{17}
\omega_d = -{\theta \ov 2\theta - \Om_d}-{\tilde{\Omega}_k\over 2\theta -\Omega_{d}},
\end{equation}
where $\tilde{\Omega}_k = (2-3\beta)\Om_k/3\beta$.
It is also interesting to study the behavior of the deceleration parameter defined as
\begin{eqnarray}\l{18}
q&=& -1 -{\dot{H} \over H^2},\nonumber\\
&=&- {  (3\beta -1)\Om_d  \ov 2\theta -\Om_d}+{(3\beta -2)(\theta+\Om_k)\ov 2\theta -\Om_d}.
\end{eqnarray}
Note that $\xi$ is very small and
  $\xi \in (-0.026, 0.015)$. Hence according to definition of    $\theta $, it  must be in $(-3.53, 1)$. This means that $\theta$  can accept the negative value. But,   observational data indicates  that the current expansion of Universe is accelerating. Several attempts have been made to justify this accelerated expansion and  this constraint requires   $\omega_d < -{1 / 3 }$ and $q <0$. Therefore, from (\ref{18}) one can see that, this model can explain the positive accelerating  expansion of Universe if $\theta > \Om_d/2$. Then the physical value of $\theta$ is in $\Om_d/2 < \theta <1$ interval.

   Also,  the equation of motion of GDE can be obtained by using (\ref{15}) and (\ref{16}) as
\begin{equation}\l{19}
{\Omega'_d}{2\theta - \Om_d \ov \Om_d  }=   3\beta[\theta +\tilde{\Om}_k] -3\beta\Om_d.
\end{equation}
here prime denote derivative with respect to efolding number $x\equiv \ln a$. Integrating of (\ref{19} ) gives
\be\l{20}
 {2\theta } \ln(\Om_d) - A \ln(|k\Om_d -1|)=3\beta A x+ c,
\ee
where
$$A = \theta -\tilde{\Om}_k$$
$$k = {1 \ov\theta +\tilde{\Om}_k }$$
and $$c = {2\theta } \ln(\Om_{d_0}) - A \ln(|k\Om_{d_0} -1|)$$
where $\Om_{d_0}$
 is the present value of dimensionless energy density of DE. \\

 At this stage we consider the stability   of GDE in the context of CBD model of gravity. The main key quantity for investigating the stability of a model is the squared adiabatic sound speed which is obtained a small perturbation  in the back ground energy density. To  derive the relevant synchronous-gauge equations of
motion, the variables that characterize the fluid are
linearized about a spatially homogeneous background \cite{20s}
 \bea\l{21}
 \R(t, x)& =&\R_b(t)+\delta\R(t,x),\\
p(t, x) &=&p_b(t)+c^2_s \delta\R(t, x)\l{22}.
\eea
The subscript $b$ denotes the
spatially homogeneous background (mean) value of the corresponding quantity, and
\be\l{23}
c^2_s = {dp \ov d\R},
\ee
is the squared adiabatic sound speed of the fluid. Using energy conservation equation yields \cite{20s}
\be\l{24}
  \nabla^2\delta\R(t,x)- {1 \ov c^2_s }{\partial^2\ov {\partial t}^2}[\delta{\R}(t,x)] =0.
\ee
This equation is an ordinary wave equation.  When $c^2_s >0$ the answer of (\ref{24}) is as  $\delta\R(t,x) =\delta\R_0\exp(\pm i [ \om t + kx])$, which is an oscillatory waves and this
 indicates
a propagation mode for the density perturbations. But,  when $c^2_s <0$, in this
case  the oscillations becomes hyperbolic  and the density perturbations
will grow with time as $\delta\R(t,x) =\delta\R_0\exp(\om t \pm i kx)$.
 Thus the perturbation is  growing  and this means the background energy of the model is  instable. Therefore, as was mentioned before, the main key quantity for investigating the stability of a model is the squared adiabatic sound speed.
 So by using (\ref{23}) and EoS $p_d = \om_d\R_d$   we have
 \be \l {25}
 c^2_s = {\dot{p} \ov \dot{\R}} = \om_d + \dot{\om}_d {\R_d \ov \dot{\R}_d},
 \ee
 taking time derivative of (\ref{17}), we have
 \be\l{26}
 \dot{\om}_d {\R_d \ov \dot{\R}_d}=({\theta - \tilde{\Om}_k }){\br [}{\Om_d   \ov (2\theta - \Om_d)^2}{\bl ]},
 \ee
  and then  we have
 \be\l{27}
 c^2_s = 2 ({\theta - \tilde{\Om}_k }) {\br [}{\Om_d - \theta  \ov (2\theta - \Om_d)^2}{\bl ]},
 \ee
since $\tilde{\Om}_k <0  $, then  it is clearly seen that for $\theta < \Om_d$ the squared adiabatic sound speed is positive and then the model is stable.

\subsection{Interacting GDE }

Some observational data such as, observational of  the galaxy cluster Abell A586,   supports the interaction between DE and DM \cite{23s}. Therefore in this section we  introduce the direct interaction between DE and
CDM and study the evolution dynamics of the  model.  Therefore according to (\ref{13}) and (\ref{14}) the  conservation equations are modified as
\begin{eqnarray}
\dot{\rho}_m + 3H\beta\rho_m &=& Q,\label{28}\\
\dot{\rho_d} + 3H\rho_d\beta (1  +\omega_d)&=&-Q,\label{29}
\end{eqnarray}
 where $Q$ denotes the interaction between GDE and CDM. A generic form of $Q$ is
not available.  Three forms which are often discussed in the literature are as  $Q = 3b^2H\rho_d, 3b^2H\rho_m, 3b^2H\rho_t $, where $b^2 > 0 $.
We want investigate the model for these three kinds of interaction. So that we consider it with a more general procedure with respect to  the pervious subsection. Hence we rewrite (\ref{7a}), (\ref{28}) and (\ref{29}) as
 \begin{eqnarray}\l{30}
  \Omega_{m} + \Omega_{d}&=& \theta + \Om_k,\\
 \dot{\Omega}_{m} + {2\dot{H} \over H}\Omega_{m} + 3\beta H\Omega_{m}&=& {Q \over 3  H^2},\l{31}\\
 \dot{\Omega}_{ed} + {2\dot{H} \over H}\Omega_{ed} + 3\beta(1+\omega_d)H\Omega_{d}&=&-{Q \over 3  H^2},\l{32}
\end{eqnarray}
where $\Om_m =\R_m/3H^2$, $\Om_d = \alpha/3H$, and $\Om_k = k/3H^2a^2$. Using (30) and (31) we have
\be\l{33}
-\dot{\Omega}_{d}+ {2\dot{H} \over H}(\theta - \Omega_{d}) + 3\beta H(\theta -\Omega_{d}) = {Q \over 3  H^2} + (2-3\beta)H\Om_k,
\ee
and by putting (\ref{32}) in (\ref{33}) one can get
\be\l{34}
2\theta \dot{ H}  + 3\beta\omega_dH^2 \Omega_{d} + 3\beta H^2 \theta = (2-3\beta)H^2\Om_k.
\ee
 Note that in GDE model we have
 \be \l{35}
 H \Omega_{d} = H_0\Omega_{0},
 \ee
 where $H_0$ and $\Omega_0$ are the value of Hubble parameter  and fraction of the ghost dark energy at the present time. Expressing (\ref{33}) in terms of efolding-number $x\equiv \ln a$,  and making use  (\ref{35}) we have
   \be\l{36}
   -\Omega'_{d}{2\theta - \Omega_{d} \over \Omega_{d}}={Q \ov 3H^3}- {3\beta(\theta - \Omega_{d})} +(2-3\beta)\Om_k.
   \ee
   Here, we can obtain the equation of state parameters of the  GDE  versus $\Omega_{d}$ as
   \be\l{37}
   \om_d = - {\theta(3\beta +2\Om_Q) + (2-3\beta)\Om_k \ov 3\beta(  2\theta - \Om_d)},
   \ee
  where $\Om_Q = Q/(3\Om_{d}H^3)$ and deceleration parameter is as
    \be\l{38}
   q = - {(\Om_Q+3\beta -1)\Om_d  \ov(  2\theta - \Om_d)}+{(3\beta -2)(\Om_k+ \theta) \ov (  2\theta - \Om_d)},
   \ee
   Although $(3\beta -2)>0$, but since $(3\beta -2)(\theta + \Om_k)$ is  smaller than $(\Om_Q+3\beta -1)\Om_d$, therefore the deceleration  parameter of the GDE model in BD scenario is negative for $\theta > \Om_d/2$, and  one can find that the EoS parameter of GDE in our model, $\om_d$, can cross the phantom divide line, $\om_d =-1$   for
   \be\l{39}
   {\Om_d \ov 2} < \theta < {3\beta (\Om_d + \tilde{\Om}_k) \ov 3\beta - 2\Om_Q}.
   \ee

   Also    we can obtain the squared adiabatic sound speed of the our model as
   \bea\l{40}
   c_s^2 &=& {dp_d \ov d\rho_d} = {\Bigr [}1-\Om_d{d \over d\Om_d}{\Bigl ]}\omega_d.
   \eea
   For getting better insight we consider the interacting GDE for three different forms of $Q$ below. However for the sake of briefness, we
   will investigate the case of $Q= 3b^2H\R_d$ in detail.

\begin{figure}[t]\label{1}
\begin{minipage}[b]{1\textwidth}
\subfigure[\label{fig1a} ]{ \includegraphics[width=.45\textwidth]%
{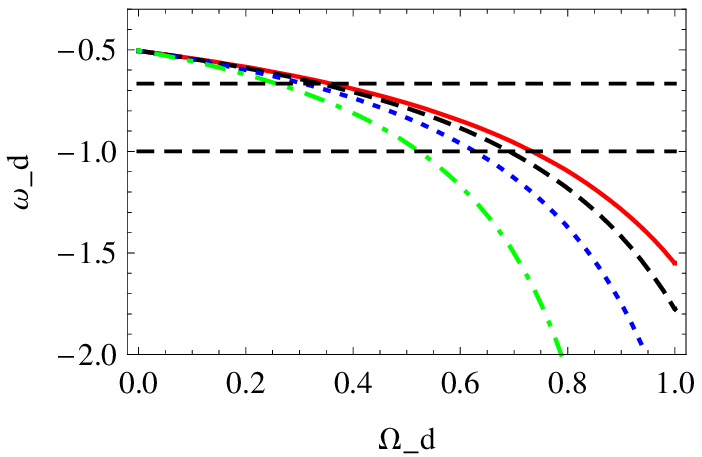}} \hspace{1cm}
\subfigure[\label{fig1b} ]{ \includegraphics[width=.45\textwidth]%
{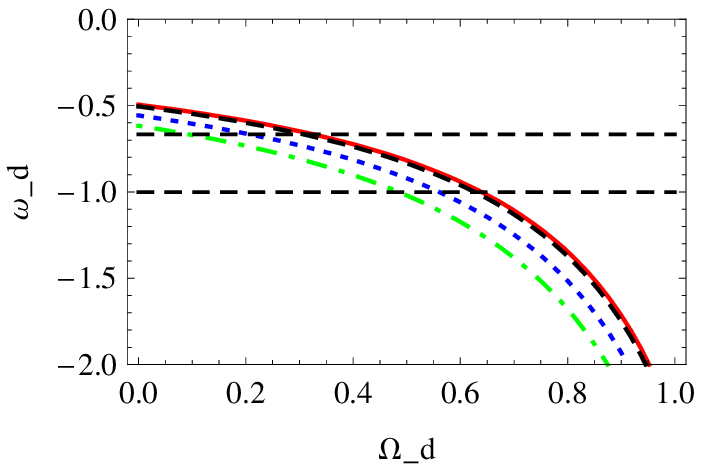}}
\end{minipage}
\caption{(a): This sub-figure shows $\om_d$ versus $\Om_d$ for $b=0.25$, $\Om_k =0.02$ and different values of $\xi$= (0.0063, {\rm red(solid)}), (0.0068, {\rm black(dashed)}), (0.0073, { blue(dotted)})$, $(0.0085, { green(dashed-dotted)}). (b): This sub-figure shows $\om_d$ versus $\Om_d$ for $\theta \simeq 0.63$, $\Om_k =0.02$ and different values of $b$=(0, { red(solid)})$, $(0.1, { black(dashed)})$, $(0.25, { blue(dotted)}), (0.45, { green(dashed-dotted)}). We have taken $Q =3b^2 H \rho_d$.  }
\end{figure}
\begin{figure}[t]\label{2}
\begin{minipage}[b]{1\textwidth}
\subfigure[\label{fig1a} ]{ \includegraphics[width=.45\textwidth]%
{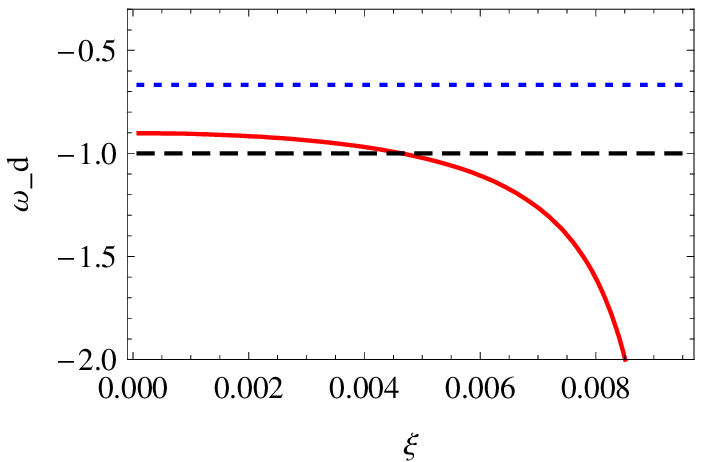}} \hspace{1cm}
\subfigure[\label{fig1b} ]{ \includegraphics[width=.45\textwidth]%
{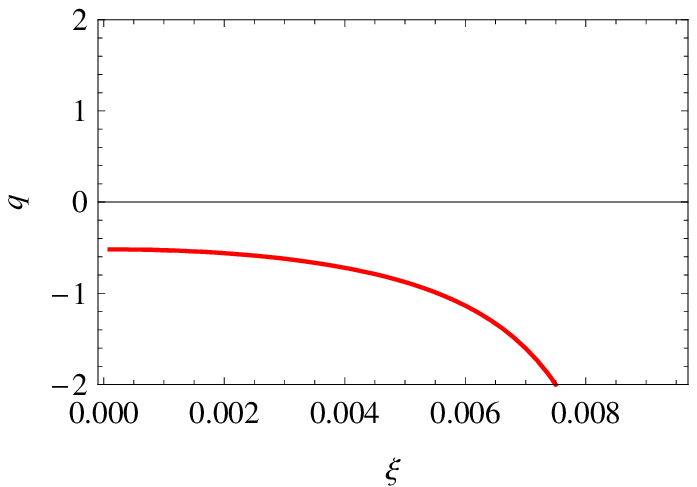}}
\end{minipage}
\caption{(a): This sub-figure shows $\om_d$ versus $\xi$ for $b=0.25$, $\Om_k=0.02$ and $\Om_d =0.76$. (b):This sub-figure shows $q$ versus $\xi$ for $b=0.25$, $\Om_k 0.02$ and $\Om_d=0.76$. We have taken $Q =3b^2 H \rho_d$.  }
\end{figure}

\begin{figure}[t]\label{3}
\begin{minipage}[b]{1\textwidth}
\subfigure[\label{fig1a} ]{ \includegraphics[width=.45\textwidth]%
{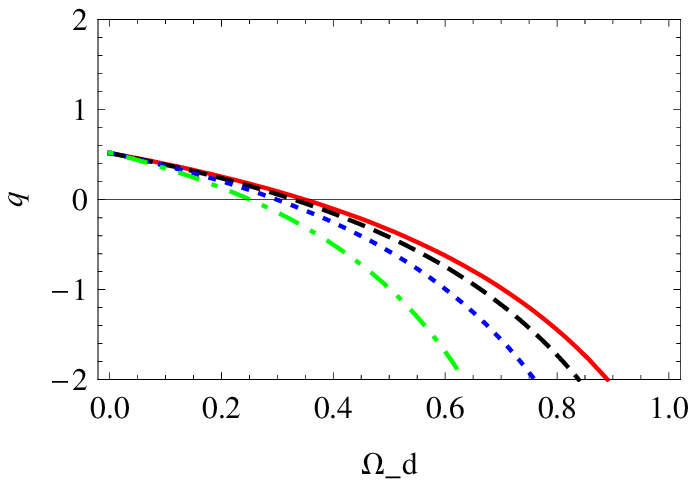}} \hspace{1cm}
\subfigure[\label{fig1b} ]{ \includegraphics[width=.45\textwidth]%
{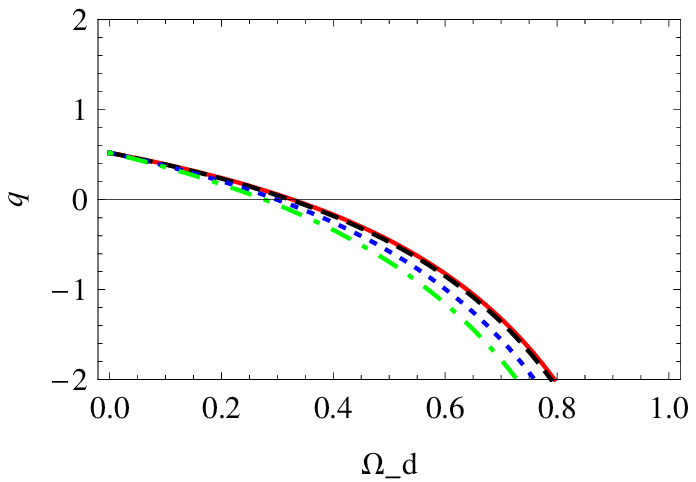}}
\end{minipage}
\caption{(a): This sub-figureshows $q$ versus $\Om_d$ for $b=0.25$, $\Om_k =0.02$ and different values of $\xi$= (0.0063, {\rm red(solid)}), (0.0068, {\rm black(dashed)}), (0.0073, { blue(dotted)})$, $(0.0085, { green(dashed-dotted)}). (b):This sub-figure shows $\om_d$ versus $\Om_d$ for $\theta \simeq 0.63$, $\Om_k =0.02$ and different values of $b$=(0, { red(solid)})$, $(0.1, { black(dashed)})$, $(0.25, { blue(dotted)}), (0.45, { green(dashed-dotted)}). We have taken $Q =3b^2 H \rho_d$.  }
\end{figure}
\begin{figure}[t]\label{4}
\begin{minipage}[b]{1\textwidth}
\subfigure[\label{fig1a} ]{ \includegraphics[width=.45\textwidth]%
{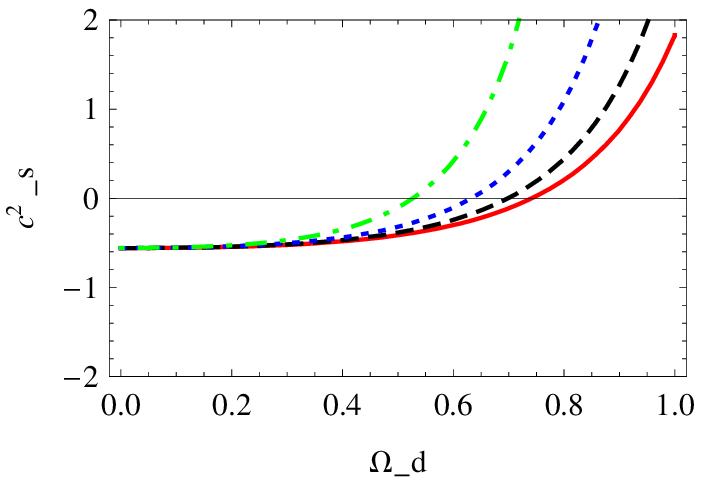}} \hspace{1cm}
\subfigure[\label{fig1b} ]{ \includegraphics[width=.45\textwidth]%
{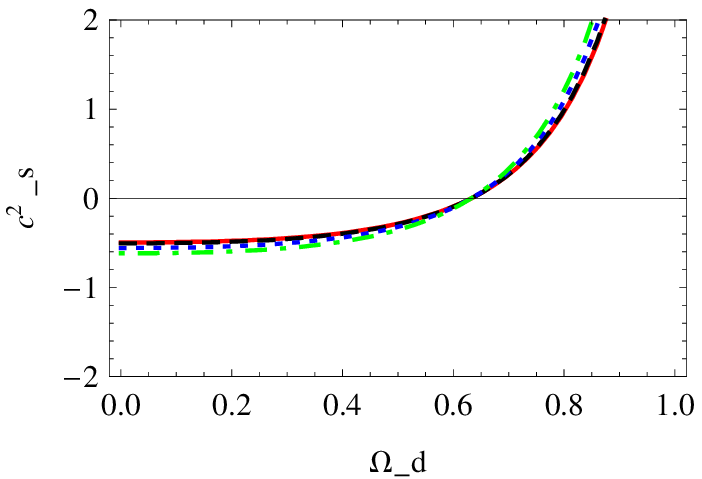}}
\end{minipage}
\caption{(a): This sub-figure shows $c^2_s$ versus $\Om_d$ for $b=0.25$, $\Om_k =0.02$ and different values of $\xi$= (0.0063, {\rm red(solid)}), (0.0068, {\rm black(dashed)}), (0.0073, { blue(dotted)})$, $(0.0085, { green(dashed-dotted)}). (b): This sub-figure shows $\om_d$ versus $\Om_d$ for $\theta \simeq 0.63$, $\Om_k =0.02$ and different values of $b$=(0, { red(solid)})$, $(0.1, { black(dashed)})$, $(0.25, { blue(dotted)}), (0.45, { green(dashed-dotted)}). We have taken $Q =3b^2 H \rho_d$. }
\end{figure}
\begin{figure}[t]\label{5}
\begin{minipage}[b]{1\textwidth}
\subfigure[\label{fig1a} ]{ \includegraphics[width=.45\textwidth]%
{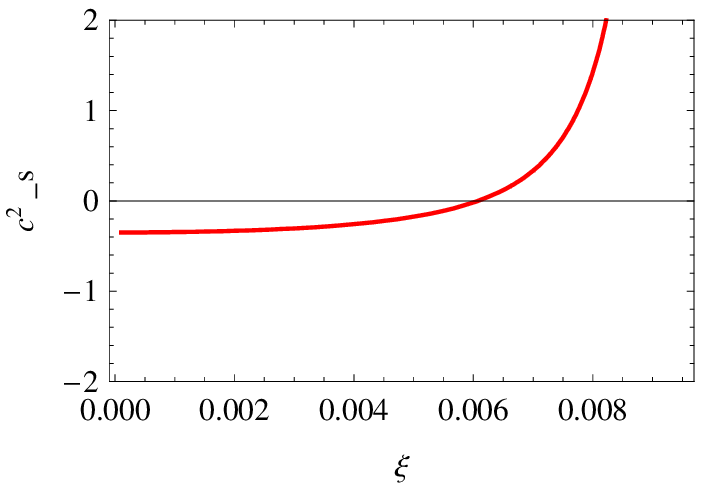}} \hspace{1cm}
\subfigure[\label{fig1b} ]{ \includegraphics[width=.45\textwidth]%
{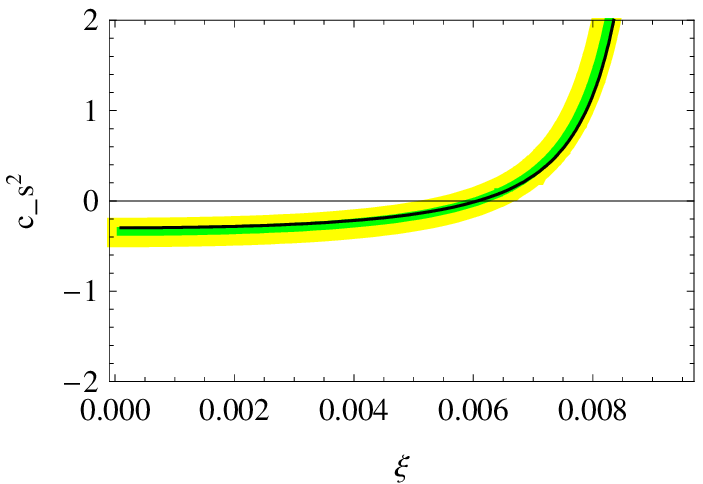}}
\end{minipage}
\caption{(a):This sub-figure shows $c^2_s$ versus $\xi$ for $b=0.25$, $\Om_k= 0.02$ and $\Om_d = 0.76$ in the case of $Q =3b^3H\rho_d$. (b): This sub-figure shows $c^2_s$ versus $\xi$ for $b=0.25$, $\Om_k 0.02$ and $\Om_d = 0.76$ in the case of $Q =3b^3H\rho_d$(\rm{yellow}), $Q =3b^3H\rho_m$(\rm{green}), $Q =3b^3H\rho_t$(\rm{black}).  }
\end{figure}

\subsubsection{$Q = 3b^2H\rho_d$}

  In this case $\Om_Q = 3b^2 $ and equations (\ref{37}) and (\ref{38})  reduce to
  \bea\l{41}
  \om_d &=& - {\theta\ov (2\theta - \Om_d)} - {\tilde{\Om}_k \ov  (2\theta - \Om_d)}-{2\theta b^2 \ov \beta( 2\theta - \Om_d)},\\
   q &=& - {(3b^2+3\beta -1)\Om_d  \ov(  2\theta - \Om_d)} +{(3\beta -2)(\Om_k+\theta) \ov (  2\theta - \Om_d)},\l{42}
     \eea
   Although $\tilde{\Om}_k <0 $ but the magnitude of it is smaller than other terms, therefore (\ref{41}) shows that $\om_d $  is always negative.     We plot the EoS parameter  versus $\Om_d$ in figure 1. Figure 1.a shows the behavior of $\om_d$ versus $\Om_d$ for $b=0.25$, $\Om_k=0.02$ and different values of $\xi$= (0.0063, {\rm red(solid)}), (0.0068, {\rm black(dashed)}), (0.0073, { blue(dotted)})$, $(0.0085, { green(dashed-dotted)}). Note that  $\theta \in (0.52, 0.74)$ for this values of $\xi$. It is clearly seen that $\om_d$ cross the line $\om_d = -1/3$  at nearly $\Om_d = 0.35$ and also cross the phantom divide line ($\om_d=-1$) at almost $\Om_d \in (0.58, 0.77)$ in which the crossing point is different via the magnitude of $\xi$. Figure 1.b indicates $\om_d$ versus $\Om_d$ for $\theta \simeq 0.63(\xi = 0.0073)$, $\Om_k=0.02$ and different values of $b$. It is seen that for $b=0$ (red-solid) the evolution of Universe  inter to the positive accelerating expansion at nearly $\Om_d\sim 0.36$ and cross the phantom line at $\Om_d\sim 0.64$. This figure shows that due to the direct interaction between DE and matter the phantom line crossing take place earlier. Also we plot the EoS parameter and deceleration parameter of GDE  versus $\xi$ in figure2. This figure shows that for all $\xi \in (0, 0.0094)$ the evolution of the Universe  completely is in positive  accelerating expansion phase and $\om_d$ can cross the line $\om_d =-1$. \\
   Also we plot the deceleration parameter, $q$, versus $\Om_d$ in Fihure3. The figure 3.a shows the behavior of deceleration  parameter for $b=0.25$, $\Om_k =0.02$ and different values of  $\xi$. According to this figure, the evolution of Universe inter to the accelerating expansion phase at nearly $\Om_d  \simeq 0.35$ for all value of $\xi\in(0.0025, 0.0094)$. Figure3b, shows $q$ versus $\Om_d$ for $\theta \simeq 0.63$  $(\xi= 0.0073)$, $\Om_k=0.02$ and different values of $b$. This figure indicates that the evolution of Universe inter to the positive accelerating expansion for $\Om_d \simeq0.3 $ for large interaction between GDE and matter which is not agree with the stability of the model. But the evolution process of  Universe inter to the accelerating phase at nearly $\Om_d \simeq 0.37 $ for $b=0$ (without interaction) and this is in a good agreement with stability of model, because the stability is satisfied only when  $\Om_d/2 <\theta  <\Om_d$.

   In this  case the adiabatic squared sound speed is
   \be
    c_s^2= 2{\Bigr [}\theta +\tilde{\Om}_k +\f{2\theta}{\beta}b^2{\Bigl]}{\Om_d-\theta \ov (2\theta -\Om_d)}\l{43}.
    \ee
    In this equation $\theta, \beta$ and also $b^2$ are positive and although $\tilde{\Om}_k$ is negative, but the value of it is very smaller than the other terms, so the squared sound speed is  positive only for $\Om_d/2 <\theta  <\Om_d$. This means that this interacting version of model can be stable. We plot the adiabatic squared sound speed of GDE versus $\Om_d$ in figure 4. This figure explicitly show that for some values of $\Om_d$, especially at the present, the squared sound speed of GDE of CBD model of gravity is positive.  Figure 4a indicate $c^2_s$ against $\Om_d$ for $b=0.25$, $\Om_k= 0.02$  and different values of $\xi$ and shows that $c^2_s$ can be positive for $\Om_d \gtrsim 0.55$. Figure 4b presents $c^2_s$ versus $\Om_d$ for $\theta \simeq 0.63 (\xi = 0.0073)$, $\Om_k = 0.02$ and different values of $b$  for the first forme of interaction, $Q =3b^2H\rho_d$. This figure tells us that a direct interaction between GDE and matter,  such as $Q= 3b^2H\rho_d$, in the cotext of CBD model of gravity has not any effect on the stability of the model. Because for all value of $b$, even $b=0$ (without interaction), the adiabatic squared sound speed has  the similar behavior versus $\Om_d$ and $c^2_s >0 $ at $\Om_d \simeq 0.63$.  This means that the GDE of CBD model is stable at the present time. The relation $c^2_s$ for an interacting case of GDE with matter is plotted versus $\xi$ in figure 5.  This figure shows that for interval $\xi\in (0.0061, 0.0094)$ which is equivalent to $\theta \in (0.42, 0.745)$ the adiabatic squared sound speed of the model is positive and then the model is stable in this form of interaction.  \\
Figure 5 present $c^2_s$ versus $\xi$. $c^2_s$ is plotted   for $Q = 3b^2H\R_d$ in figure 5a and for all three forms of $Q$ in figure 5b. These two sub-figures  show that the behavior of $c^2_s$ versus $\xi$ is similar for all three forms of interactions.\\
    Finally the equation (\ref{36}) becomes
   \be\l{44}
   \Omega'_{d}{2\theta - \Omega_{d} \over \Omega_{d}}=  3\beta(\theta -\tilde{\Om}_k) -3(b^2+\beta) \Omega_{d}.
   \ee
   its analytical solution reads
   \be\l{45}
   2\theta\ln\Om_d - \Om_1 \ln(|1-k_1\Om_d|) = 3\beta(\theta -\tilde{\Om}_k)x +c,
   \ee
   where
\bea\l{46}
\Om_1 &=&{\theta(\beta +2b^2) +\beta\tilde{\Om}_k \ov \beta +b^2},\\
k_1&=& {\beta +b^2 \ov \beta(\theta -\tilde{\Om}_k)}.\l{47}
\eea
and $c=  2\theta\ln\Om_{d0} - \Om_1 \ln(|1-k_1\Om_{d0}|) $ is the integration constant and $\Om_{d0}$ is the fraction of dark energy  at present time. The relation of $\Om_d$ versus $x=\ln(a)$ is shown in Figure 6. From the figure6 we can see  that $\Om_d$
 varies from 0 at early time to 1 at late time.

\begin{figure}[ht]\label{6}
\centerline{ \includegraphics[width=7cm] {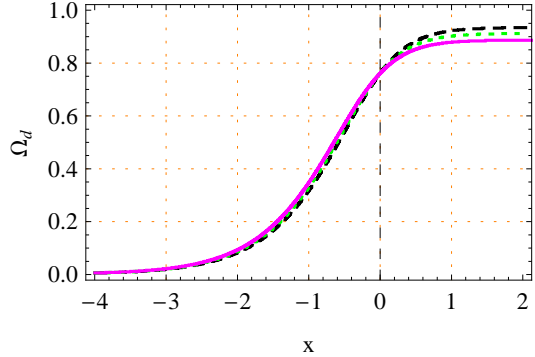}},
\caption{ This figure show $\Om_d$ versus $x=\ln(a)$ for   $b=0.1$, $\om_k =0.02$  and different values of $\xi$ = (0.0065, black(dashed)), (0.0075, green(dotted)), (0.0085, red(solid)). We have taken $Q = 3b^2H\R_d$.}
\end{figure}

\subsubsection{ $Q=3b^2H\rho_m$}
In this case $\Om_Q = 3b^2(\theta +\Om_k -\Om_d )/\Om_d $ and equation (\ref{37}), (\ref{38}) and (\ref{40}) reduce to
 \bea\l{48}
  \om_d &=& - {(\theta+\tilde{\Om}_k)\ov(  2\theta - \Om_d)}+{2\theta b^2 \ov \beta(  2\theta - \Om_d)} -{2\theta b^2(\theta +\Om_k) \ov \beta\Om_d(  2\theta - \Om_d)},\\
   q &=& -{(3\beta-3b^2-1)\Om_d \ov (  2\theta - \Om_d)}+{(3\beta-3b^2 -2)(\theta + \Om_k)\ov(  2\theta - \Om_d) } ,\l{49}\\
   c_s^2&=& 2\om_1{\Om_d-\theta \ov (2\theta -\Om_d)^2} + \om_2{3\Om_d -4\theta \ov \Om_d(2\theta -\Om_d)^2},
   \eea
   where
   $$\om_1 =\theta +\tilde{\Om}_k -{2\theta \ov \beta}b^2, $$
   $$\om_2 = {2\ov \beta}\theta b^2 (\theta +\Om_k).$$
   From the above equations are  seen that when $\Om_d \rightarrow 0$, $\om_d$, $q$, $c^2_s$ tend to $-b^2(\theta + {\Om}_k)/\beta\Om_d$, $(3\beta -b^2 -2)(\theta + {\Om}_k)/2\theta$, $ -\om_2/\theta\Om_d $  respectively. We see that at the early time $\om_d$ and $c^2_s$ are divergent.\\
The equation of motion for $\Om_d$ is
\be\l{51}
   \Omega'_{d}{2\theta - \Omega_{d} \over \Omega_{d}}= 3{\Big [}(\beta-b^2)\theta -  ({2\ov 3}-\beta +b^2)\Om_k{\Big ]} -3(\beta -b^2)\Omega_{d}.
   \ee
Its analytical solution gives
 \be\l{45}
  2\theta\ln\Om_d - \Om_2 \ln(|1-k_2\Om_d|) = {\Big [}3(\beta- b^2)\theta -(3b^2+2-3\beta){\Om}_k{\Big ]}x +c,
   \ee
   where
\bea\l{46}
\Om_2 &=&{(\beta - b^2)\theta +(b^2-\beta +2/3)\Om_k\ov \beta -b^2},\\
k_2&=& {\beta -b^2 \ov (\beta - b^2)\theta -(b^2-\beta +2/3)\Om_k}.\l{47}
\eea
and $c=  2\theta\ln\Om_{d0} - \Om_2 \ln(|1-k_2\Om_{d0}|) $ is the integration constant.

\begin{figure}[t]\label{7}
\begin{minipage}[b]{1\textwidth}
\subfigure[\label{fig1a} ]{ \includegraphics[width=.45\textwidth]%
{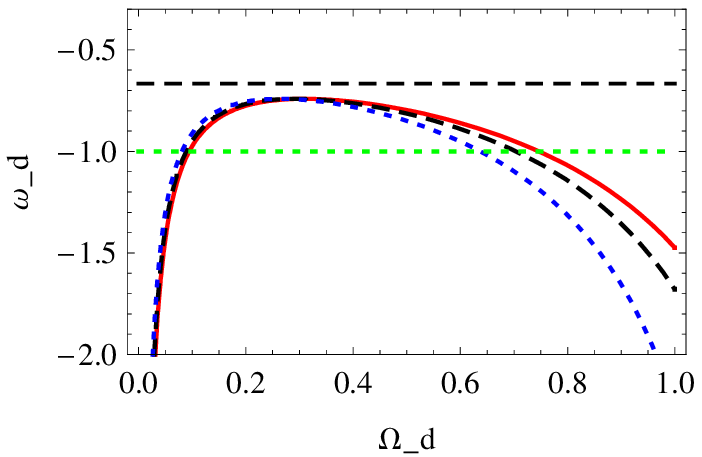}} \hspace{1cm}
\subfigure[\label{fig1b} ]{ \includegraphics[width=.45\textwidth]%
{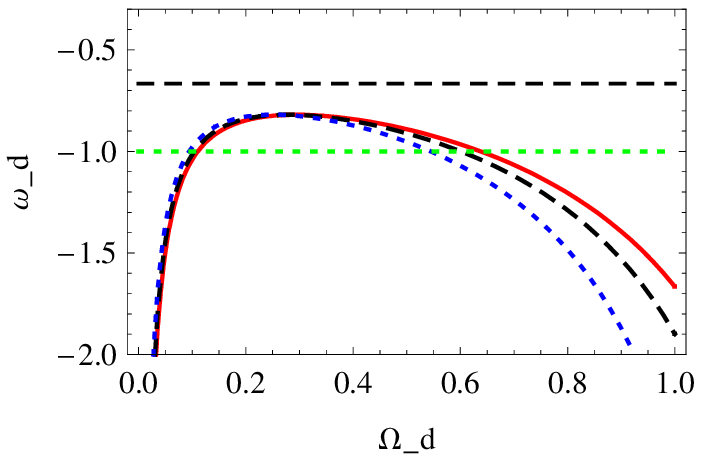}}
\end{minipage}
\caption{(a): This sub-figure shows $\om_d$ versus $\Om_d$ for $b=0.25$, $\Om_k=0.02$ and different values of $\xi= 0.0063$ ({\rm red-solid}), 0.0068 ({\rm black-dashed}), 0.0073 ({\rm blue-dotted}) in the case of $Q =3b^2H\rho_m$. (b): This sub-figure shows $\om_d$ versus $\Om_d$ for $b=0.25$, $\Om_k=0.02$ and different values of $\xi= 0.0063$ ({\rm red-solid}), 0.0068({\rm black-dashed}), 0.0073( {\rm blue-dotted}) in the case of $Q =3b^2H\rho_t$.   }
\end{figure}
\begin{figure}[t]\label{8}
\begin{minipage}[b]{1\textwidth}
\subfigure[\label{fig1a} ]{ \includegraphics[width=.45\textwidth]%
{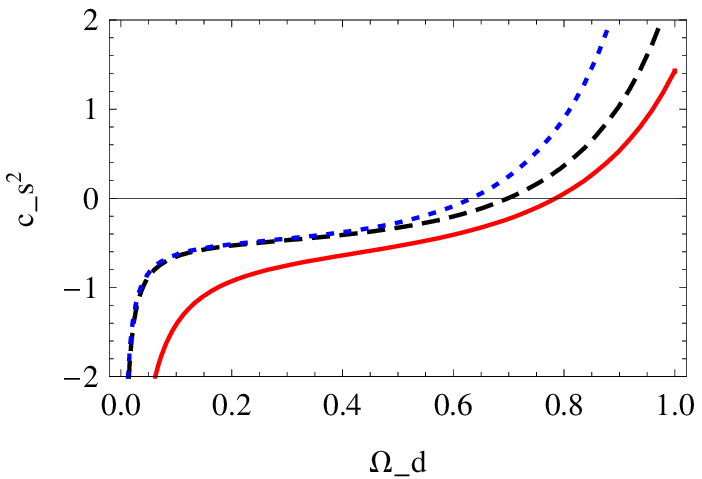}} \hspace{1cm}
\subfigure[\label{fig1b} ]{ \includegraphics[width=.45\textwidth]%
{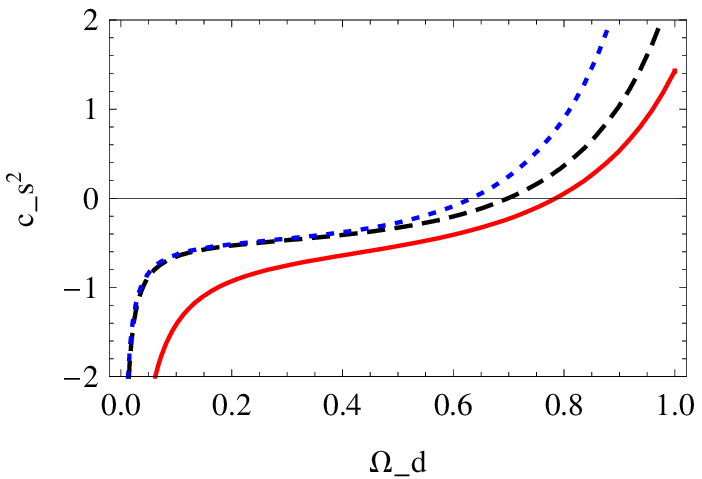}}
\end{minipage}
\caption{(a): This sub-figure shows $c^2_s$ versus $\Om_d$ for $b=0.25$, $\Om_k=0.02$ and different values of $\xi= 0.0063$ ({\rm red-solid}), 0.0068 ({\rm black-dashed}), 0.0073 ({\rm blue-dotted}) in the case of $Q =3b^2H\rho_m$. (b): This sub-figure shows $c^2_s$ versus $\Om_d$ for $b=0.25$, $\Om_k=0.02$ and different values of $\xi= 0.0063$ ({\rm red-solid}), 0.0068({\rm black-dashed}), 0.0073( {\rm blue-dotted}) in the case of $Q =3b^2H\rho_t$.    }
\end{figure}

 \subsubsection{ $Q=3b^2H\rho_t$}
In this case $\Om_Q = 3b^2(\theta +\Om_k )/\Om_d $ and equation(\ref{37}), (\ref{38}) and (\ref{40}) reduce to
\bea
   \om_d &=& - {(\theta+\tilde{\Om}_k)\ov(  2\theta - \Om_d)} -{2\theta b^2(\theta +\Om_k) \ov \beta\Om_d(  2\theta - \Om_d)},\\
   q &=& -{(3\beta-1)\Om_d \ov (  2\theta - \Om_d)}+{(3\beta    -3b^2-2)(\theta + \Om_k) \ov (  2\theta - \Om_d)},\\
  c_s^2&=& 2\om_1{\Om_d-\theta \ov (2\theta -\Om_d)^2} + \om_2{3\Om_d -4\theta \ov \Om_d(2\theta -\Om_d)^2},
   \eea
   where
   $$\om_1 =\theta +\tilde{\Om}_k,  $$
   $$\om_2 = {2\ov \beta}\theta b^2 (\theta +\Om_k).$$
   It is seen that the quantity $\om_d$, $q$, $c^2_s$ have the same behavior which we obtain for $Q=3b^2H\R_m$. This means that $\om_d$ and $c^2_s$ are divergent in the early time.\\

Finally in this case the equation of motion for dimensionless density energy parameter is
\be\l{51}
   \Omega'_{d}{2\theta - \Omega_{d} \over \Omega_{d}}= 3{\Big [}(\beta-b^2)\theta -  ({2\ov 3}-\beta +b^2)\Om_k{\Big ]} -3\beta \Omega_{d}.
   \ee
Integration (\ref{44}), gives
 \be\l{45}
  2\theta\ln\Om_d - \Om_3 \ln(|1-k_3\Om_d|) = {\Big [}3(\beta- b^2)\theta -(3b^2+2-3\beta){\Om}_k{\Big ]}x +c_3,
   \ee
   where
\bea\l{46}
\Om_3 &=&{(\beta - b^2)\theta +(b^2-\beta +2/3)\Om_k\ov \beta },\\
k_3&=& {\beta  \ov (\beta - b^2)\theta -(b^2-\beta +2/3)\Om_k}.\l{47}
\eea
and
$$c_3=  2\theta\ln\Om_{d0} - \Om_3 \ln(|1-k_3\Om_{d0}|). $$

We plot $\om_d$ versus $\Om_d$ for $Q = 3b^2H\R_m$ and $Q = 3b^2H\R_t$ in figure 7. These two sub-figures show  that the behavior of $\om_d$ versus $\Om_d$ is similar for two different kinds of interaction and also they show  the evolution of Universe is completely  in positive  accelerating expansion phase,  because $\om_d$ is always less than $-1/3$. These figures indicate   $\om_d$  will increase from $-\infty$ to a local maximum  below $\om_d = -1/3$  and then decrease to the below of $\om_d =-1$. This means that $\om_d$ cross the line $\om_d =-1$  two times. One time from phantom phase to quintessence phase nearly at early time and another time  from quintessence phase to phantom phase in which at the present ($\Om_d = 0.76)$ the second crossing has took place.

 Figures 8(a) and 8(b) show the squared sound speed versus $\Om_d$ for $Q = 3b^2H\R_m$ and $Q = 3b^2H\R_t$ respectively. These figures indicate $c_s^2$ is negative in $ 0< \Om_d <0.63$ for these two kinds of interaction,   and for $\Om_d > 0.63$ one can obtain a positive value for $c_s^2$ by a suitable choice of $\xi$ parameter. This means that for these two kinds of interaction the GDE model is stable in the present time  in the context of CBD model of gravity.
%======================= Section 4 (Conclusion) ======================
%=====================================================================
\section{Conclusion}
Recently, Ghost dark energy is introduced \cite{32}. This kind of dark energy is  a phenomenological vacuum  energy which is rooted from the Veneziano ghost of QCD.  The ghost   dark energy model is proportional to Hubble parameter.  According to results of \cite{35}, the adiabatic squared sound speed of the GDE model in the context of  Einsteinian  theory of gravity  is negative and then this model is instable. And also it is shown that this model has a  behavior like a cosmological constant at the late time ($\omega =-1$) and the equation of state can never  cross $-1$.
We study the stability and the evolution of this model in the context of Brans-Dicke model which has non-minimal coupling with matter, namely chameleon Brans-Dicke (GDECBD) model.
 At first, we investigated the GDE of  CBD model without any interaction between GDE and CDM in a non flat FLRW metric. In this investigation we obtain the EoS parameter, deceleration parameter, the equation of motion for GDE fraction parameter and the adiabatic squared sound speed of the model. The obtained quantities show that  the GDE model in the context of CBD scenario can describe the positive accelerating expansion phase of Universe and the EoS parameter of dark energy can cross the phantom divide line, also  the squared sound speed of the DE is positive, for a suitable choice if $\xi$ parameter. We studied the
cosmological dynamics of the model by considering three usual forms of  interaction between
DE and CDM. Considering the evolution of dimensionless density energy parameter, we have shown that $\Om_d$ varies from 0 at early time to 1 at late time.  We fund that the evolution of Universe in GDE model of CBD scenario is completely in positive accelerating expansion phase for three forms of interaction. Also  we obtained  $\om_d$ crosses the line $\om_d=-1$ from phantom phase to quintessence phase  for $Q=3b^2H\R_d$ case, but the behavior of $\om_d$ versus $\Om_d$ is similar for $Q=3b^2H\R_m$ and $Q=3b^2H\R_t$ and also      $\om_d$ cross the line $\om_d =-1$  two times. One time from phantom phase to quintessence phase nearly at early time and another time  from quintessence phase to phantom phase in which at the present ($\Om_d = 0.76)$ the second crossing has took place.

 Finally we found the squared sound speed of GDE model in CBD framework  for  three kinds of interaction in the FRW Universe. All of our calculations show that    $c_s^2$ is negative in $ 0< \Om_d <0.63$   and for $\Om_d > 0.63$ one can obtain a positive value for $c_s^2$ by a suitable choice of $\xi$ parameter. This means that the  interacting and non-interacting case of   GDE model is stable in the present time  in the context of CBD model of gravity.

 %=====================================================================
%======================= Reference ===================================
%=====================================================================
\section{Aknowledgement}
 The work of  Kh. Saaidi have been supported financially  by  University of Kurdistan, Sanandaj, Iran, and  he would like thank to the University of Kurdistan for supporting him in sabbatical period.

\end{document}